\documentclass{pasj00}

\begin{document}
\SetRunningHead{T. Kato et al.}{On the Possibility of V4444 Sgr as a Recurrent Nova}

\Received{}
\Accepted{}

\title{On the Possibility of V4444 Sgr as a Recurrent Nova}

\author{Taichi \textsc{Kato}}
\affil{Department of Astronomy, Kyoto University,
       Sakyo-ku, Kyoto 606-8502}
\email{tkato@kusastro.kyoto-u.ac.jp}

\author{Hitoshi \textsc{Yamaoka}}
\affil{Faculty of Science, Kyushu University, Fukuoka 810-8560}
\email{yamaoka@rc.kyushu-u.ac.jp}

\email{\rm{and}}

\author{Seiichiro \textsc{Kiyota}}
\affil{Variable Star Observers League in Japan (VSOLJ),
       1-401-810 Azuma, Tsukuba, Ibaraki 305-0031}
\email{skiyota@nias.affrc.go.jp}



\maketitle

\begin{abstract}
   V4444 Sgr (Nova Sgr 1999) has been recently suggested to be a possible
recurrent nova based on the detection of the preexisting dust, which may
have been formed in a previous outburst.  We examined this possibility
using the available outburst observations, including the VSNET
observations.  We also noticed that this nova was recorded in the
OGLE II public photometric database, which covered both preoutburst and
late decline phases.  The outburst light curve constructed from these
observations showed a plateau phase, particularly in the OGLE II
$I$-band light curve, which was followed by a more rapid decline.
The light curve resembled those of ``fast'' recurrent
novae and candidates in the presence of the distinct plateau phase.
The outburst spectrum and the preoutburst magnitude more resembled those
of classical novae, but they do not exclude the possibility of a
recurrent nova.
\end{abstract}

\section{Introduction}

   Our observational knowledge of recurrent novae has recently made
a significant advancement, particularly led by the most recent detections,
and prompt announcements through VSNET, of three recurrent nova outbursts:
U Sco in 1999 (\cite{sch99uscoiauc}; \cite{mun99usco}; \cite{kah99uscoSSS};
\cite{lep99usco}; \cite{anu00usco}; \cite{eva01uscoIR};
\cite{iij02uscospec}),
CI Aql in 2000 (\cite{tak00ciaql}; \cite{kis01ciaql}; \cite{mat01ciaql};
\cite{mat03ciaql}; \cite{bur01ciaql}),
and IM Nor in 2002 (\cite{lil02imnoriauc7789}; \cite{kat02imnor}).
Theoretical models have been able to reproduce the observed features
of the recurrent novae (\cite{hac00uscoburst}, \cite{hac00uscoqui}).
Even a prediction of supersoft X-ray phase and turnoff has been made
\citep{hac01ciaql}.  However, the most recent two recurrent novae,
CI Aql and IM Nor, have a significant departure from the classical
descriptions of recurrent novae (\cite{web87RN1}; \cite{web87RN2};
\cite{anu92RN}), in their classical nova-like light curves and spectra.
\citet{kat02imnor} even proposed that IM Nor and CI Aql comprise
a new subclass of recurrent novae with massive ejecta and
long recurrence times, which may be an indication that these objects
are transitional objects between classical novae and recurrent novae.
In any case, recurrent novae are now recognized to comprise a more
heterogeneous population of novae than was previously thought.

   V4444 Sgr (Nova Sgr 1999) was discovered by M. Yamamoto
\citep{kus99v4444sgriauc}.  The nova nature was confirmed by the presence
of strong, broad emission lines
(\cite{lil99v4444sgriauc}; \cite{gar99v4444sgriauc}).
The nova declined very rapidly, at a rate corresponding to $t_2 \sim$ 3.5 d
\citep{kaw00v4444sgr}, which qualifies V4444 Sgr to be a very fast nova
according to the classification by \citet{GalacticNovae}.

   \citet{kaw00v4444sgr} reported, from spectropolarimetric observations,
that the nova showed intrinsic polarization, whose properties were explained
by scattering by surrounding small dust grains.  \citet{kaw00v4444sgr}
further suggested that these dust grains may have produced by a previous
explosion.  \citet{ven02v4444sgrIR} performed near-infrared spectroscopy,
and detected thermal emissions which can be attributed to dust.
\citet{ven02v4444sgrIR} discussed, from the lack of a detectable
dust-forming episode in the visual light curve, that the dust is likely
to be preexisting.  From these observations, \citet{ven02v4444sgrIR}
suggested that V4444 Sgr may be a recurrent nova.

   We recently noticed that V4444 Sgr was detected as a transient-type
variable star (OGLE II BUL-SC19-V1874) during the Optical Gravitational
Lensing Experiment II (OGLE II; \cite{OGLE2}) bulge variable star
database \citep{woz02OGLE2var}.
Since this observation covered the important stages of the
outburst, we reexamined the outburst light curve using this archival data,
and discuss the possibility of a recurrent nova in view of the recent
knowledge of recurrent novae.

\begin{figure*}
  \begin{center}
    \FigureFile(78mm,78mm){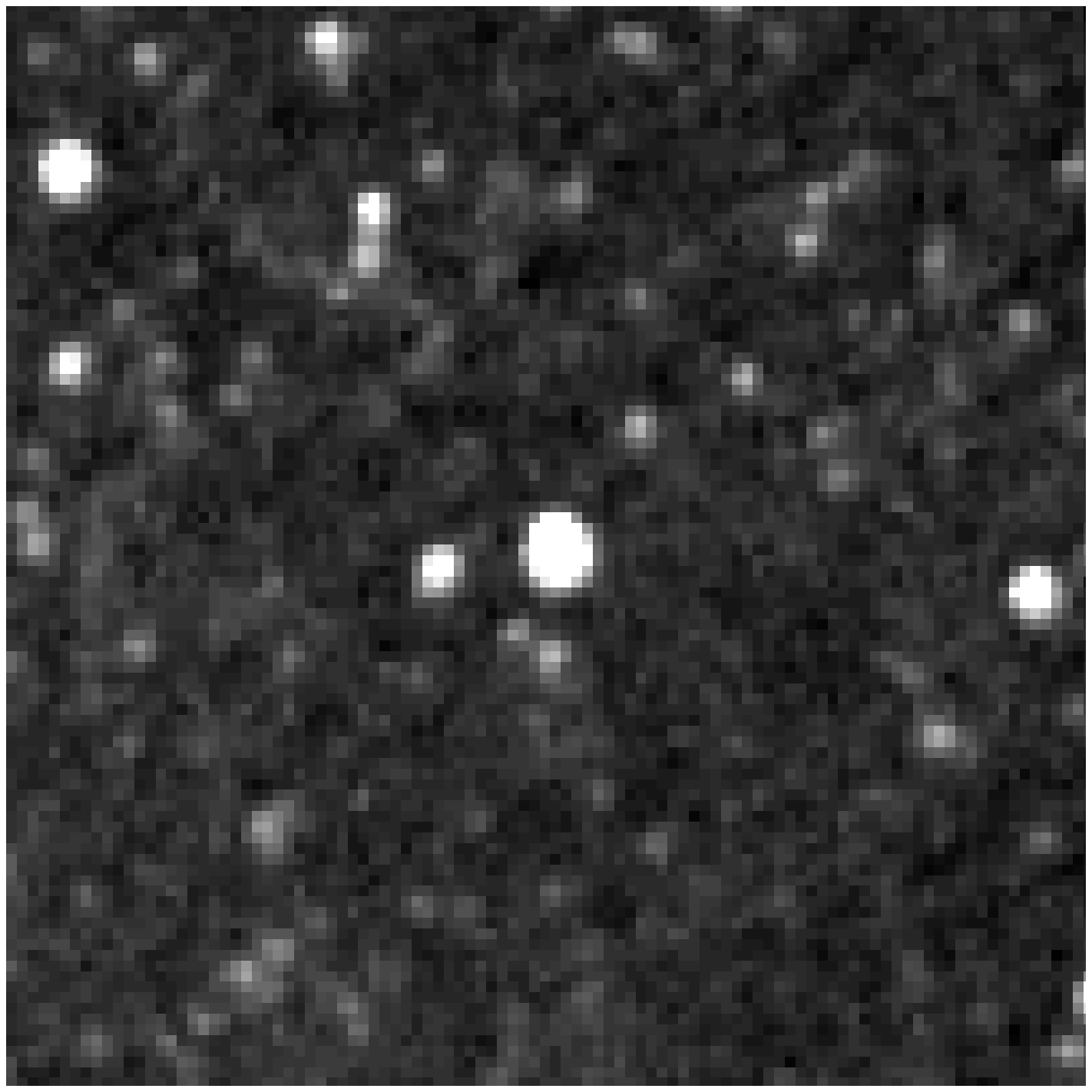}
    \hskip 2mm
    \FigureFile(78mm,78mm){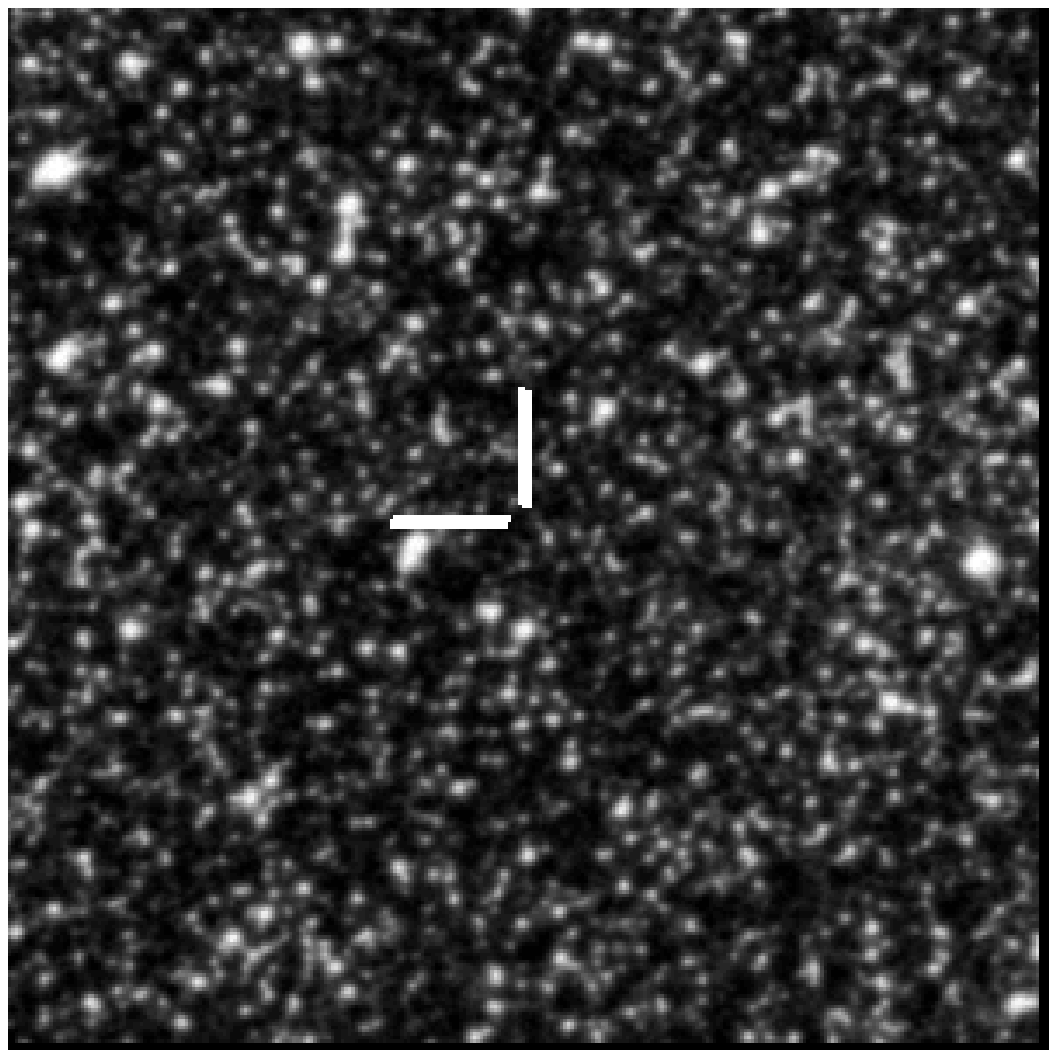}
  \end{center}
  \caption{Identification of V4444 Sgr.  North is up, east is left,
           5 arcmin square.
  Left: Outburst image of V4444 Sgr.  Right: The quiescent
  position (tick marks) on the DSS 2 red image.
  See figure \ref{fig:id2} for an enlargement of DSS 2.}
  \label{fig:id}
\end{figure*}

\section{Observation and Data Analysis}

   The visual, photographic and CCD data used to construct the overall
outburst light curve are from the reports to the VSNET \citep{VSNET}.

   The multicolor CCD observations were made by SK, with a 25-cm
Schmidt--Cassegrain telescope and an AP-7 CCD.  The magnitudes were
determined using the neighboring Tycho stars.  The data are listed
in table \ref{tab:multicolor}.

   The newly analyzed data are from the OGLE II variable star public
archive.\footnote{
$\langle$ftp://bulge.princeton.edu/ogle/ogle2/bulge\_dia\_variables/\\plain\_text/BUL\_SC19/bul\_sc19\_1874.dat$\rangle$.}
In the OGLE II original data, the nova was below the limit of detection
until HJD 2451290.898 (1999 April 21).  The OGLE II PSF photometry
yielded unsuccessful fitting results on 19 nights between
HJD 2451295.682 (1999 April 26) and HJD 2451342.749 (1999 June 13).
This period corresponds to a period when the nova was brighter than
$V$ = 13, which must have caused a saturation on OGLE II images.
We used the $I$-band magnitudes after this period.

\begin{table}
\caption{CCD multicolor photometry.}\label{tab:multicolor}
\begin{center}
\begin{tabular}{ccccc}
\hline\hline
HJD$-$2450000 &  $B$  &  $V$  & $R_{\rm c}$ \\
\hline
1298.154 &  9.83 &  8.99 &  8.29 \\
1299.187 &  9.94 &  9.21 &  8.40 \\
1305.166 & 11.03 &  ---  &  9.23 \\
1306.153 & 11.09 & 10.64 &  9.24 \\
1307.185 & 11.32 & 10.86 &  9.34 \\
1309.153 & 11.53: & 11.13 & 9.45 \\
1311.195 & 11.75 & 11.31 &  9.60 \\
1319.144 & 12.46 & 11.92 & 10.04 \\
1330.081 & 12.78 & 12.73 & 10.89 \\
1331.186 &  ---  & 12.74 &  ---  \\
1342.161 & 13.89 & 13.90 & 11.54 \\
1343.147 & 13.92 & 13.69 & 11.69 \\
1367.197 & 14.56 & 14.34 & 12.62 \\
1369.081 & 14.82 & 14.30 & 12.72 \\
1384.170 & 14.92 &  ---  & 12.91 \\
1409.010 &  ---  & 14.93 & 13.56 \\
1426.989 &  ---  & 15.13 & 14.40 \\
\hline
\end{tabular}
\end{center}
\end{table}

\section{Astrometry and Pre-outburst Magnitude}\label{sec:id}

   \citet{kus99v4444sgriauc} reported that nothing obvious appears at
this location of the nova on the Digital Sky Survey (DSS).  We tried
to make refined astrometry, in order to obtain a more stringent limit
to the pre-outburst magnitude.

   Astrometry of the outbursting V4444 Sgr was performed on a CCD
image taken by SK (figure \ref{fig:id}).  A measurement (UCAC1 system,
115 reference stars; fitting error was $\sim$\timeform{0''.2})
has yielded a position of \timeform{18h 07m 36s.198},
\timeform{-27D 20' 12''.95} (J2000.0).
The published and newly determined positions are summarized in
Table \ref{tab:astrometry}.  Our position basically agrees with the 
OGLE-II position.  The small difference is possibly caused by 
the difference in the selection of reference catalogs.

\begin{figure}
  \begin{center}
    \FigureFile(88mm,88mm){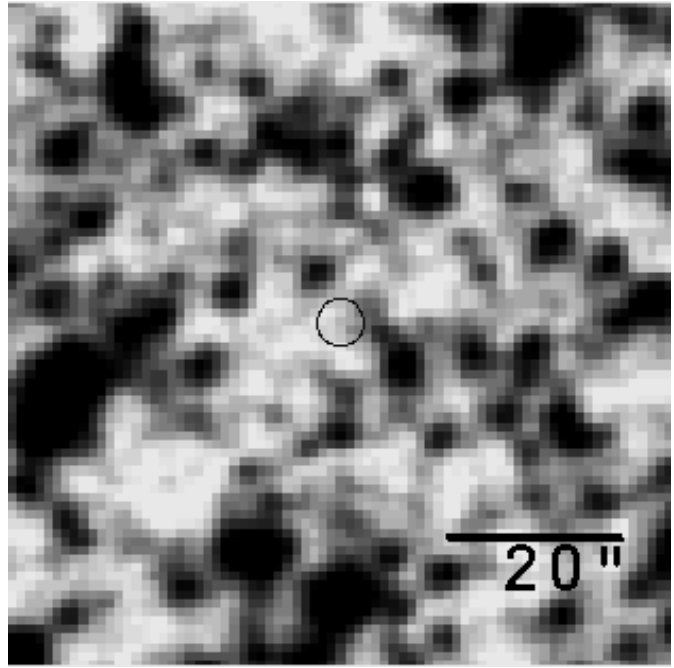}
  \end{center}
  \caption{Enlargement of the DSS 2 red image.  The central
  circle (5 arcseconds in diameter) represents the position of V4444 Sgr.}
  \label{fig:id2}
\end{figure}

   On DSS 2 red images, a faint object (red magnitude about 19) exists
within  \timeform{0''.6} of our position (figure \ref{fig:id2}).
This object can be the nova in quiescence, and constrains the lower
limit of the outburst amplitude to be 11 mag.

\begin{table}
\begin{center}
\caption{Astrometry of V4444 Sgr.}\label{tab:astrometry}
\begin{tabular}{lcc}
\hline\hline
Source    & R.A. & Decl. \\
          & \multicolumn{2}{c}{(J2000.0)} \\
\hline
\citet{kus99v4444sgriauc}
                   & 18 07 36.22 & $-$27 20 13.5 \\
OGLE II catalog
                   & 18 07 36.22 & $-$27 20 13.4 \\
This work
                   & 18 07 36.198 & $-$27 20 12.95 \\
\hline
\end{tabular}
\end{center}
\end{table}

\section{Outburst Light Curve}\label{sec:lc}

\begin{figure}
    \FigureFile(88mm,60mm){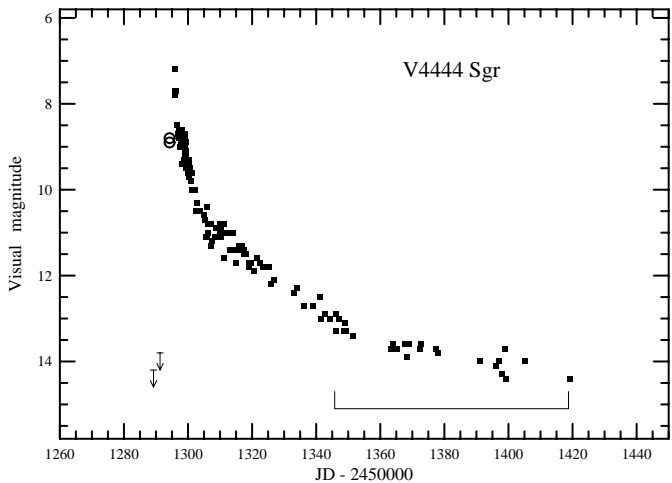}
  \caption{Overall visual outburst light curve of V4444 Sgr.  A $V$-band
  maximum observation by \citet{zis99v4444sgriauc7154}
  has been supplemented.
  The open circles and the arrows are the discovery photographic
  observations and prediscovery upper limits, respectively.
  The labeled epoch
  correspond to the OGLE II $I$-band light curve in figure \ref{fig:ogle}
  (the OGLE II data are not plotted on this figure).}
  \label{fig:overall}
\end{figure}

\begin{figure}
  \begin{center}
    \FigureFile(88mm,60mm){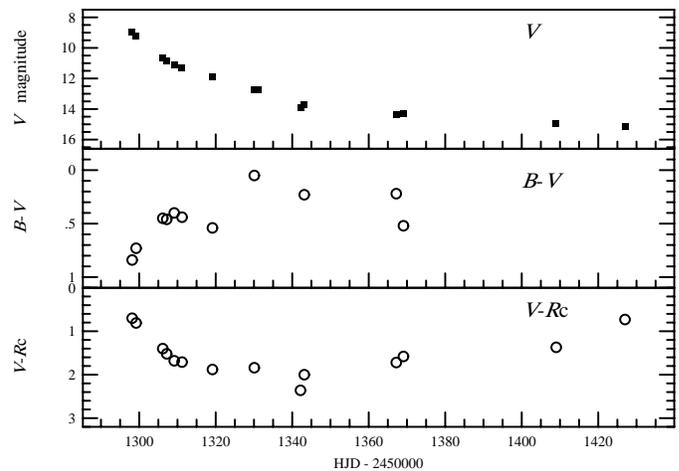}
  \end{center}
  \caption{Multicolor photometry of V4444 Sgr.}
  \label{fig:multi}
\end{figure}

   The overall light curve constructed from the reports to VSNET is
presented in figure \ref{fig:overall}.  CCD multicolor photometry is
shown in figure \ref{fig:multi}.  The light curve seems to be
composed of a rapidly fading initial stage and a stage of a more
gradual decline.

\begin{figure}
  \begin{center}
    \FigureFile(88mm,60mm){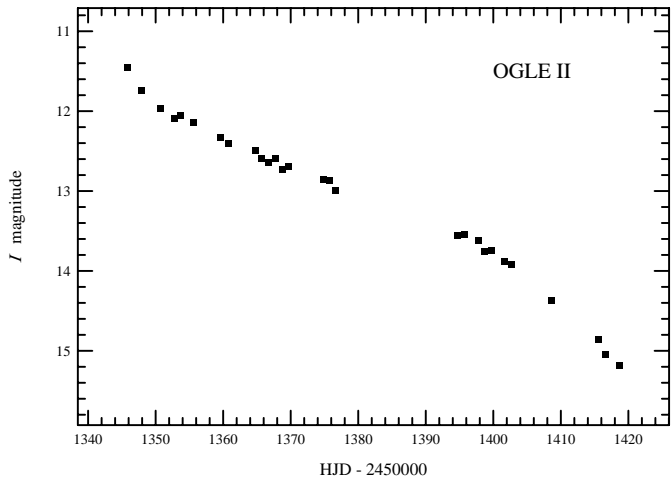}
  \end{center}
  \caption{OGLE II $I$-band observations.  The observed epoch correspond
  to the slowly declining stage of the outburst.  The error bars are
  smaller than the marks.}
  \label{fig:ogle}
\end{figure}

\section{Pre-outburst Light Curve}

\begin{figure}
  \begin{center}
    \FigureFile(88mm,60mm){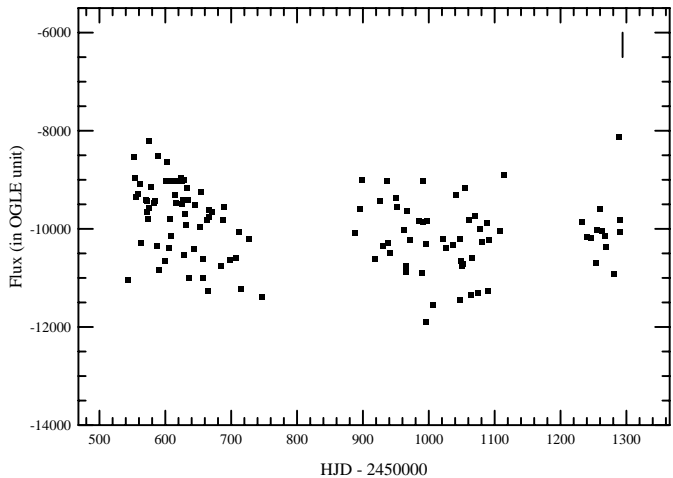}
  \end{center}
  \caption{OGLE II $I$-band pre-outburst observations.  The raw fluxes
  (in OGLE II unit) are shown, which have a large zero-point offset
  (not corrected here).
  There is no indication of a brightening before the outburst.
  The vertical tick at HJD 2451294.232 corresponds to the epoch of
  the initial detection by M. Yamamoto.
  }
  \label{fig:pre}
\end{figure}

   Figure \ref{fig:pre} shows OGLE II $I$-band pre-outburst observations.
The raw fluxes are shown, which has a large zero-point offset
(approximately $-$10000, not corrected here; the OGLE II unit 1000
approximately corresponds to $I$ = 17.2).
There is no indication of a brightening before the
outburst.  Since the initial detection ($m_{\rm pg}$ = 8.8) by Yamamoto
was made on HJD 2451294.235, this observation severely constrains the
rise time to be less than 3.3 d.  These pre-outburst observations
preclude the possibility of a steady rising for $\sim$1000 d in
V533 Her and LV Vul \citep{rob75novapreeruption}, or a month-long
premaximum rise in V1500 Cyg \citep{kuk75v1500cyg2}.

\section{V4444 Sgr as a Recurrent Nova Candidate}

   We discuss the possibility of a recurrent nova in view of the recent
knowledge of recurrent novae.

\subsection{Spectroscopic Features}

   The spectra of V4444 Sgr taken during the outburst \citep{kaw00v4444sgr}
are indistinguishable from those of Fe\textsc{ii}-class classical novae
\citep{wil92novaspec}, and are unlike those of U Sco
(\cite{bar79usco}; \cite{sek88usco}).
Such Fe\textsc{ii}-class spectra have
historically been recorded only in classical novae.  The recent discoveries
of Fe\textsc{ii}-class recurrent novae (CI Aql and IM Nor), however,
makes this distinction less meaningful than in the past.

\subsection{Light Curve}

   The light curve bears some resemblance to those of recurrent novae.
As shown in figure \ref{fig:ogle}, the $I$-band light curve shows an almost
linear decline between JD 2451350 and JD 2451400.\footnote{
   The tendency of the light curve may look slightly different from the
   visual observation (figure \ref{fig:overall}).  This is probably because
   of the contamination from emission lines and because of the difficulty
   in visually measuring the faint object $V \sim$ 14.  Since the $I$-band
   is less affected by strong emission lines, the $I$-band light curve
   provides the best estimate of the continuum variation.
}  This stage with a linear decline seems to corresponds to the
{\it plateau phase}, which is characteristic to recurrent novae
(\cite{hac00uscoburst}; \cite{hac01ciaql}; \cite{hac02v2487ophproc}).
During the plateau phase, continuum emission from the accretion disk
significantly contributes to the overall light (see \cite{sek88usco};
see also \cite{hac00uscoburst} for a theoretical estimate).  The relative
increase of the contribution from the continuum may explain the decrease
of the $V-R_{\rm c}$ color index during this period
(cf. figure \ref{fig:multi}).

   The observed duration of the plateau phase ($\sim$ 50 d) is longer
than $\sim$16 d of the U Sco outburst in 1999 \citep{mat03usco} and
$\sim$20 d for the suggested recurrent nova V2487 Oph
\citep{hac02v2487ophproc}, but is shorter than $\sim$1.4--1.7 yr of
the CI Aql outburst in 2000
\citep{mat03ciaql}.\footnote{
  There was a dip-like fading during the late stage of the 2000 outburst
  of CI Aql.  The duration of the plateau phase before this dip was
  $\sim$160 d \citep{mat01ciaql}.
}

   After JD 2451400, the nova
started to fade more rapidly, which is consistent with the termination
of the plateau phase, rather than a more continuous decrease in the
decline rate in most of classical novae (cf. \cite{ClassicalNovae}).
The lack of reddening during
the fading stage seems to preclude a possibility of a dust-forming episode
[see \citet{due81novatype} for representative color changes].
The characteristics of the light curve of V4444 Sgr thus resembles
those of ``fast'' recurrent novae.  If V4444 Sgr indeed turns out
to be a recurrent nova, V4444 Sgr is expected to have intermediate
properties between U Sco and CI Aql.

\subsection{Progenitor}

   As shown in figure \ref{sec:id}, the progenitor of V4444 Sgr was
about mag 19.  If we adopt the maximum $M_V$ of $-6.4\pm0.4$ as in the
recurrent nova U Sco (\cite{hac00uscoburst}; \cite{hac00uscoqui}),
the preoutburst $M_V$ is estimated to be $\sim$ $+$4.5.
This value is a typical prenova magnitude of a classical nova
\citep{war86NLabsmag}, but looks slightly faint for a recurrent nova,
since recurrent nova outbursts would require a high accretion rate
\citep{hac00uscoqui}.
If the maximum $M_V$ is $-9.0\pm0.2$ \citep{kaw00v4444sgr} assuming
the relation in classical novae \citep{dellaval95novaabsmag},
the preoutburst magnitude of $M_V \sim$ 2 is, alternatively, slightly
too bright for a classical nova \citep{war86NLabsmag}.

   If V4444 Sgr is indeed a recurrent nova, we may need an obscuration
by the previously ejected materials, as has been proposed to explain
the faint quiescence of U Sco \citep{hac00uscoqui}.  Considering
the likely existence of dust before the outburst, this possibility
is not surprisingly unlikely.  It is also known that the recurrent nova
IM Nor has a faint quiescence \citep{kat02imnor}, which can be attributed
to the short-period and probably eclipsing nature of this system
(cf. \cite{wou03imnor}); this possibility would be tested by future
observations.

\section{Summary}

   V4444 Sgr (Nova Sgr 1999) has been recently suggested to be a possible
recurrent nova based on the detection of the preexisting dust, which may
have been formed in a previous outburst.  We examined this possibility
using the available outburst observations.  We have newly identified
the nova in the OGLE II bulge variable star database.  The OGLE II
record covered both preoutburst and late decline phases of this nova.
The outburst light curve showed a plateau phase, particularly in the
OGLE II $I$-band light curve, which was followed by a more rapid decline.
Quasi-simultaneous optical multicolor photometry precludes the possibility
of a dust-forming episode as the origin of this fading.
The light curve thus resembled those of ``fast'' recurrent
novae and candidates in the presence of the distinct plateau phase.
The outburst spectrum and the preoutburst magnitude more resembled those
of classical novae, but they do not exclude the possibility of a
recurrent nova.  If V4444 Sgr is an indeed a recurrent nova, the relatively
faint pre-outburst magnitude would require an obscuration by the preexisting
dust prior to the 1999 outburst, or the object may be similar to
the short-period, probably eclipsing recurrent nova, IM Nor.
We have been able to exclude the possibility of a pre-outburst gradual
rise observed in a few past classical novae.

\vskip 3mm

The authors are grateful to the OGLE team in making their variable star
data publicly available.
The authors are grateful to many observers who reported observations
to VSNET.
This work is partly supported by a grant-in aid [13640239, 15037205 (TK),
14740131 (HY)] from the Japanese Ministry of Education, Culture, Sports,
Science and Technology.
This research has made use of the Digitized Sky Survey producted by STScI, 
the ESO Skycat tool, the VizieR catalogue access tool.

\end{document}